%% file: main.tex
\documentclass[lettersize,journal]{IEEEtran}
\usepackage{amsmath,amsfonts}
\usepackage{algorithmic}
\usepackage{algorithm}
\usepackage{array}
\usepackage{textcomp}
\usepackage{stfloats}
\usepackage{url}
\usepackage{verbatim}
\usepackage{graphicx}
\usepackage{cite}
\usepackage{amsmath}
\usepackage{utfsym}
\usepackage{makecell}
\usepackage{multirow}
\usepackage{booktabs}
\usepackage{caption}
\usepackage{subcaption}
\usepackage{dcolumn}
\usepackage{hyperref}
\usepackage{bm}
\usepackage{threeparttable}

\usepackage[table]{xcolor}
\usepackage{colortbl}

\newcolumntype{P}[1]{>{\centering\arraybackslash}p{#1}}

\begin{document}

\renewcommand{\arraystretch}{1.3} 

\title{A Study of the Removability of Speaker-Adversarial Perturbations}

\author{Liping Chen$^{*}$,~\IEEEmembership{Senior Member,~IEEE,} Chenyang Guo$^{*}$, Kong Aik Lee,~\IEEEmembership{Senior Member,~IEEE,} \\ Zhen-Hua Ling,~\IEEEmembership{Senior Member,~IEEE,} and Wu Guo,~\IEEEmembership{Member,~IEEE,}
\thanks{This paper is an extension of our published conference paper \cite{guo2024removal}.}
\thanks{Liping Chen, Chenyang Guo, Zhen-Hua Ling and Wu Guo are with the University of Science and Technology of China, China (e-mail: lipchen@ustc.edu.cn, chenyangguo@mail.ustc.edu.cn, zhling@ustc.edu.cn, guowu@ustc.edu.cn). Kong Aik Lee is with The Hong Kong Polytechnic University, Hong Kong SAR, China (e-mail: kong-aik.lee@polyu.edu.hk).}
\thanks{$^{*}$ contributed equally to this paper.}
\thanks{\emph{Corresponding author: Liping Chen}.}
\thanks{This work was supported in part by the National Key Research and Development Program Project 2024YFE0217200, the National Natural Science Foundation of China under Grant U23B2053, and the Fundamental Research Funds for the Central Universities WK2100000043.}}

\markboth{Journal of \LaTeX\ Class Files,~Vol.~14, No.~8, August~2021}%
{Shell \MakeLowercase{\textit{et al.}}: A Sample Article Using IEEEtran.cls for IEEE Journals}


\maketitle

\begin{abstract}

Recent advancements in adversarial attacks have demonstrated their effectiveness in misleading speaker recognition models, making wrong predictions about speaker identities. On the other hand, defense techniques against speaker-adversarial attacks focus on reducing the effects of speaker-adversarial perturbations on speaker attribute extraction. These techniques do not seek to fully remove the perturbations and restore the original speech. To this end, this paper studies the removability of speaker-adversarial perturbations. Specifically, the investigation is conducted assuming various degrees of awareness of the perturbation generator across three scenarios: ignorant, semi-informed, and well-informed. Besides, we consider both the optimization-based and feedforward perturbation generation methods. Experiments conducted on the LibriSpeech dataset demonstrated that: 1) in the ignorant scenario, speaker-adversarial perturbations cannot be eliminated, although their impact on speaker attribute extraction is reduced, 2) in the semi-informed scenario, the speaker-adversarial perturbations cannot be fully removed, while those generated by the feedforward model can be considerably reduced, and 3) in the well-informed scenario, speaker-adversarial perturbations are nearly eliminated, allowing for the restoration of the original speech. Audio samples can be found in \url{https://voiceprivacy.github.io/Perturbation-Generation-Removal/}.

\end{abstract}


\section{Introduction}

\IEEEPARstart{O}{ver} the past decades, speaker recognition techniques have gone through remarkable advancements. Specifically, the introduction of i-vector \cite{dehak2010front} enables the representation of speaker characteristics within variable-length utterances as fixed-length embedding vectors. The rapid advancement of neural networks (NNs) in recent years has further promoted speaker modeling capabilities through x-vector \cite{snyder2017deep} and its derivatives, such as ECAPA-TDNN \cite{desplanques2020ecapa} and ResNet \cite{zeinali2019but}. On the other hand, neural network models have been found to be vulnerable to adversarial perturbations in input samples \cite{2014Explaining}, prompting investigations into adversarial attack targeting speaker recognition models, including NN-based \cite{zhang2023imperceptible, abdullah2021hear, 9053076, li2020universal} and the conventional i-vector \cite{9053076}.
Furthermore, the ability of adversarial speech to cheat identity recognition enables its application in voice privacy protection, aimed at protecting speaker attributes from being correctly extracted \cite{v-cloak,chen2023voicecloak,UniAP,chen2024adversarial}.


As speaker attributes play a vital role in speaker recognition, adversarial attack targeting speaker recognition systems pose significant risks. For instance, a targeted attack on a specific speaker verification system misleads the system into misclassifying an input speech as that of the targeted speaker, potentially facilitating malicious impersonation of the target. Such risks have driven the development of defense techniques against speaker-adversarial speech. In this context, pre-processing methods aimed at removing the impacts of adversarial perturbations have been investigated. Notably, methods have been developed that operate without prior knowledge of the speaker-adversarial speech, employing strategies like vector quantization\cite{characterinzing}, compression\cite{chen2022towards}, smoothing\cite{characterinzing}, adding noise\cite{a15080293}, filtering\cite{9053643}, generative models \cite{TIFS.2021.3116438,bai2023diffusion}, among others. Moreover, denoising methods have been investigated for removing speaker-adversarial perturbations in scenarios where adversarial speech is either accessible or not, enabling denoising models to be trained using pairs of adversarial-original\cite{zhang20g_interspeech,li2023unified} and noisy-original\cite{chang2021defending} speech utterances, respectively.

While reducing the impacts of speaker-adversarial perturbations in speaker recognition tasks, current research on defense techniques does not seek to fully restore the original speech signal. To this end, this paper studies the removability of speaker-adversarial perturbations, with the aim to 1) restore the original speech, 2) reduce the effects on speaker attributes extraction, and 3) reduce the impacts on speech utility. Particularly, the elimination of perturbation is measured with the scale-invariant signal-to-noise ratio (SI-SNR) and mean square error (MSE) computed between the processed and original utterances. The impact on speaker attribute extraction is assessed with automatic speaker verification (ASV) evaluations. The utility concerns the speech quality of the processed speech and the preservation of speaker-independent information, such as linguistic content and prosody, which are evaluated with automatic speech recognition (ASR) and pitch extraction tasks, respectively.


Moreover, the removability of speaker-adversarial perturbations is studied in three scenarios based on the knowledge available to the remover namely, \emph{ignorant}, \emph{semi-informed}, and \emph{well-informed}. The ignorant scenario assumes that the remover is lack of knowledge of the specific speaker-adversarial speech it processes. In the semi-informed scenario, the remover is assumed to have access to the speaker-adversarial speech generation algorithm, enabling its training using pairs of original and adversarial speech utterances. The well-informed scenario assumes that the perturbation generator and remover are mutually informed of each other. Particularly, in this paper, the generator and remover are jointly trained to establish their information sharing.


This study investigates two types of speaker-adversarial perturbation generation methods: the optimization-based method and the feedforward method. In the former, the momentum iterative fast gradient step method (MI-FGSM) \cite{dong2018boosting} and the gradient relevance attack (GRA) \cite{zhu2023boosting} are examined. In the latter, the symmetric saliency-based encoder-decoder (SSED) structure \cite{yao2023symmetric} is utilized. Experiments were conducted on the LibriSpeech dataset \cite{panayotov2015librispeech} to evaluate the efficacy of the perturbation removers in eliminating speaker-adversarial perturbations, thereby restoring the original speech in the ignorant, semi-informed, and well-informed scenarios, respectively. Our experimental results demonstrate that the effects of speaker-adversarial perturbations on speaker attribute extraction can be reduced in all three scenarios. Additionally, the removability of the speaker-adversarial perturbations in each scenario are observed as follows:

1) In the ignorant scenario, the perturbations cannot be eliminated, failing to restore the original speech. Besides, adverse effects on speech utility were observed.

2) In the semi-informed scenario, the perturbations generated by the optimization-based MI-FGSM and GRA methods can hardly be reduced. However, those generated by the feedforward SSED model can be larged removed, resulting in considerable restoration of the original speech.

3) In the well-informed scenario, the speaker-adversarial perturbations are nearly eliminated, thereby restoring the original speech.

The contributions of this study are summarized as follows:

1) We conduct a comprehensive investigation into the removability of the speaker-adversarial perturbations from speech utterances. We define the criteria for perturbation removal and the scenarios based on the awareness of the adversarial speech by the remover.

2) We conducted extensive experiments to evaluate the removability of speaker-adversarial perturbations generated with the optimization-based and feedforward methods in the ignorant, semi-informed, and well-informed scenarios. The experimental findings are anticipated to enhance the understanding of speaker-adversarial perturbation removal and the restoration of the original speech signal.

\section{Background}
This section presents the background of our work. First, the speaker extractor is described. Subsequently, the methods for generating speaker-adversarial perturbations are briefly outlined, including the optimization-based and feedforward methods. Thereafter, the existing speaker-adversarial perturbation removal methods examined in this study are briefly outlined, including non-informed defense methods and denoising techniques.

\subsection{Speaker encoder}
A speaker encoder extracts speaker attributes from speech utterances and represents them as embedding vectors. The utterances of the same speaker are close to each other in the embedding vector space. Denoting an utterance as $\mathcal{O}_{1:T}$ where $T$ is the number of frames, the speaker embedding extraction process can be described as:

\begin{equation}
    \label{eq. speaker embedding}
    {\mathcal{F}}_{\theta}:~{\mathcal O}_{1:T}\mapsto{\bm \phi}.
\end{equation}
In (\ref{eq. speaker embedding}), $\mathcal{F}_{\theta}$ is  the \emph{speaker encoder} that maps $\mathcal O_{1:T}$ into a fixed-length embedding vector $\bm \phi$. Here, $\theta$ is the parameter set of the mapping function. The vector $\bm \phi$ captures the long-term vocal characteristics of the speaker within the utterance and is referred to as \emph{speaker embedding}.

Regarding whether the speaker attributes are modeled in a generative or discriminative manner, $\phi$ can be classified into two broad categories, namely, (i) \emph{generative} embedding, and (ii) \emph{discriminative} embedding. The i-vector \cite{dehak2010front} is a typical example of the former. The x-vector \cite{snyder2017deep} and its variants \cite{8683760,9053209, Villalba2019StateoftheArtSR,desplanques2020ecapa,zeinali2019but, 9463712} are representatives of the latter, with ECAPA-TDNN \cite{desplanques2020ecapa} and ResNet \cite{zeinali2019but} being the most widely architectures.

\subsection{Optimization-based methods}
Assume an original speech utterance ${\mathcal{O}}$ and a speaker extractor function $\boldsymbol{y}=\mathcal{F}_{\theta}\left( {{\mathcal{O}}} \right)$, where $\theta$ is the model parameters and ${\boldsymbol{y}}$ is the extracted speaker embedding. Define the loss function for speaker-adversarial speech generation as $L\left(\boldsymbol{y}\right)$. The optimization-based methods produce adversarial perturbations in the gradient direction that increase the loss function. The adversarial samples are then obtained via adding the perturbations to the original inputs. Among the optimization-based methods, the fast gradient sign method (FGSM) \cite{2014Explaining} generates adversarial samples in a single step. Thereafter, the iterative strategy was introduced, leading to iterative FGSM (I-FGSM)\cite{kurakin2017adversarial}. Following that, the momentum mechanism was incorporated, and the momentum I-FGSM (MI-FGSM) \cite{dong2018boosting} was developed.

Specifically, with MI-FGSM, the speaker-adversarial version of ${\mathcal{O}}$ is obtained as follows:
\begin{equation}
    \label{eq. MI-FGSM 1}
    {{{\tilde{{\mathcal{O}}}}}_{i + 1}} = {\rm{Clip}}^\epsilon \left\{ {{{{\tilde{{\mathcal{O}}}}}_i} + \alpha \cdot {\rm{sign}}\left( {{{\boldsymbol{g}}_{i + 1}}} \right)} \right\},
\end{equation}
where $i=0,1,2...,I-1$ indicates the iterations with $I$ to be the number of iterations. The ${{{\tilde{{\mathcal{O}}}}}_0}$ is initialized to ${{\mathcal{O}}}$ and $0<\alpha<\epsilon$ is the step size. The ${\rm Clip}^{\epsilon}$ function limits $\tilde{{\mathcal{O}}}_{i+1}$ to the vicinity of ${\mathcal{O}}$ satisfying the $L_{\infty}$ norm bound with $\epsilon$ to be the intensity. The variable ${\boldsymbol g}_i$ accumulates the gradients of the previous $i$ iterations with a momentum decay factor ${\eta}$ as follows:
\begin{equation}
    \label{eq. MI-FGSM 2}
    {{\boldsymbol{g}}_{i + 1}} = \eta \cdot{{\boldsymbol{g}}_i} + \frac{{{\nabla _{\tilde{{\mathcal{O}}}_i}}\left( {L\left( {{\tilde{\boldsymbol{y}}_i} } \right)} \right)}}{{{{\left\| {{\nabla _{\tilde{{\mathcal{O}}}_i}}\left( {L\left( {\tilde{{\boldsymbol{y}}}_i } \right)} \right)} \right\|}_1}}},
\end{equation}
where $\tilde{{\boldsymbol{y}}}_i=\mathcal{F}_{\theta}\left({\tilde{{\mathcal{O}}_i}}\right)$. Finally, the speaker-adversarial utterance is obtained as ${\tilde{\mathcal O}}_I$.

Based on MI-FGSM, variants have been developed to enhance transferability, including GAR \cite{zhu2023boosting}, NI-FGSM \cite{linnesterov}, etc.

\subsection{SSED} 
\begin{figure}[t]
    \centering
    \label{task}
    \includegraphics[scale=0.4]{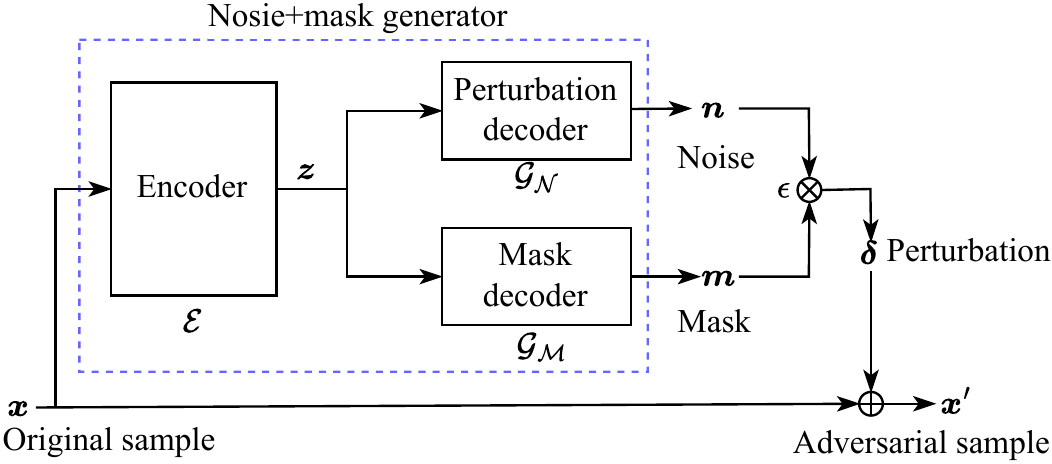}
    \caption{Architecture of the perturbation generator of SSED. The noise and mask are generated by the noise+mask generator, represented with the rectangular box of the blue dotted line.}
    \label{fig. SSED}
    \end{figure}

The symmetric saliency-based encoder-decoder (SSED) \cite{ssed} is a feedforward neural network for speaker-adversarial speech generation. As shown in Fig. \ref{fig. SSED}, it is composed of an encoder $\mathcal{E}$, a noise decoder $\mathcal{G}_\mathcal{N}$, and a saliency map decoder $\mathcal{G}_\mathcal{M}$. Given a sample from the original speech $ \boldsymbol x$, it is firstly encoded by the encoder $\mathcal{E}$ into a latent vector $ \boldsymbol z$. Then $ \boldsymbol z$ is decoded into a noise vector $ \boldsymbol n$ by $\mathcal{G}_\mathcal{N}$. Meanwhile, $\mathcal{G}_\mathcal{M}$ is applied on $ \boldsymbol z$, decoding it into the mask ${\boldsymbol m}$. Both $ \boldsymbol n$ and $ \boldsymbol m$ are in the same size as $ \boldsymbol x$. The perturbation $ \boldsymbol{\delta}$ is obtained by the element-wise product between $ \boldsymbol n$ and $ \boldsymbol m$. Finally, $ \boldsymbol{\delta}$ is added to $ \boldsymbol x$, resulting in its adversarial form as follows:
\begin{equation}
\label{eq: SSED adversarial sample}
{ \boldsymbol x}' =  \boldsymbol x + \underbrace{\epsilon\cdot(  \boldsymbol n \odot  \boldsymbol m)}_{ \boldsymbol{\delta}},
\end{equation}
where $\epsilon$ denotes the attack intensity. Specifically, the convolutional residual blocks are used in the encoder/decoder $\mathcal{E}$, $\mathcal{G}_\mathcal{N}$, and $\mathcal{G}_\mathcal{M}$. In this work, the modules applied in the process of generating the noise \( \boldsymbol{n} \) and \( \boldsymbol{m} \) from the original sample \( \boldsymbol{x} \) form the \emph{noise+mask generator}.

\subsection{Non-informed defense methods}
Non-informed defense methods for speaker-adversarial perturbations were developed under the assumption that the generation algorithm for adversarial speech is inaccessible. Some of them operate in the time domain. For example, the quantization (QT) method \cite{characterinzing} rounds the amplitude of the waveform sample to the nearest integer multiple of the factor $\lambda$, which will have an impact on adversarial perturbations. The compression strategy \cite{chen2022towards} reduces adversarial perturbations via audio signal compression methods such as OPUS\cite{vos2013voice}, SPEEX\cite{valin2016speex}, AMR\cite{781503}, AAC\cite{bosi1997iso}, MP3\cite{hacker2000mp3}, etc. Smoothing methods\cite{characterinzing} (median, average, etc.) mitigate adversarial waveform samples by smoothing within a sliding window. The adding noise method\cite{a15080293} incorporates small, random noise (typically Gaussian noise) into speaker-adversarial speech to mitigate the effects of perturbations. Some methods operate in the frequency domain, such as filtering approaches. Specifically, these methods include down-sampling \cite{characterinzing}, low-pass filtering\cite{chen2022towards}, band-pass filtering\cite{chen2022towards}, and others. Furthermore, generative models \cite{defensegan,joshi2021adversarial,Bai2023DiffusionBasedAP} have been explored for adversarial attack defense, adopting the strategy of mapping adversarial samples back to their original counterparts. In this context, architectures such as generative adversarial network (GAN) \cite{goodfellow2020generative}, variational autoencoder (VAE) \cite{kingma2013auto}, and diffusion model \cite{ho2020denoising} have been examined.

\subsection{Denoising methods}
Speech denoising techniques are developed to remove the noise from noisy speech\cite{AZARANG20201,1164453,1455809,6932438,6639038,Wang2005,7364200}. In the past decade, the introduction of neural networks into speech denoising has substantially advanced its progress. Among these, in the ideal ratio mask (IRM) method \cite{6639038}, the denoised speech is derived by multiplying the time-frequency bins of the noisy signal with their predicted IRMs. Subsequently, the complex idea ratio mask (cIRM) \cite{7364200} was introduced to enhance both the magnitude and phase responses of noisy speech. Derivatives such as DCCRN \cite{hu20g_interspeech} and FRCRN \cite{9747578} have since been developed. Between them, DCCRN achieved first place in the real-time track and second place in the non-real-time track of the Interspeech 2020 Deep Noise Suppression (DNS) challenge\cite{reddy2020the}, while FRCRN achieved second place in the real-time full-band track of the ICASSP 2022 DNS challenge\cite{dubey2022icassp}, demonstrating state-of-the-art performance on wide-band benchmark datasets. 

Notably, denoising models can be utilized to remove adversarial perturbations in both ignorant and semi-informed scenarios. In the ignorant scenario, the model is trained using pairs of original and noisy speech utterances\cite{chang2021defending}, with the noisy speech generated by adding additive noise to the original. In the semi-informed scenario, where speaker-adversarial speech is available, the model is trained with original and adversarial utterance pairs\cite{zhang20g_interspeech,li2023unified}. It maps the adversarial speech to its original version, thereby removing the adversarial perturbation.

    \begin{figure}[t]
    \centering
    \includegraphics[scale=0.85]{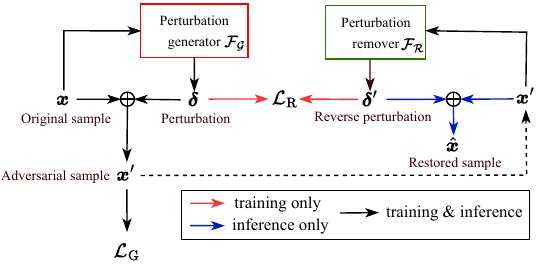}
    \caption{General joint training framework for the speaker adversarial perturbation generator and remover.}
    \label{fig: joint training framework}
    \end{figure}

\section{Introducing Joint Training for Adversarial Removal}
The joint training strategy is proposed in the well-informed scenario. Specifically, given an original speech utterance, the generator produces speaker-adversarial perturbations, obtaining the adversarial speech by incorporating these perturbations. The remover predicts the reverse of the perturbations based on the adversarial speech and restores the original form by adding them to the adversarial speech.

\subsection{Overall framework}
The overall framework of the joint training framework is depicted in Fig. \ref{fig: joint training framework}. Given a speech sample $\boldsymbol{x}$, the adversarial perturbation $\delta$ is generated by a perturbation generator ${\mathcal F}_{\mathcal G}$ and then added to $\boldsymbol{x}$, resulting in the adversarial sample $\boldsymbol{x}'$. Subsequently, $\boldsymbol{x}'$ is input to the perturbation remover ${\mathcal F}_{\mathcal R}$, which estimates the reverse of $\delta$, denoted as $\delta'$. Finally, the restored sample $\hat{\boldsymbol{x}}$ is obtained by adding $\delta'$ to $\boldsymbol{x}'$. Within this framework, both ${\mathcal F}_{\mathcal G}$ and ${\mathcal F}_{\mathcal R}$ are feedforward networks.


As illustrated in Fig. \ref{fig: joint training framework}, the speaker-adversarial speech generator is supervised by the loss function \({\mathcal{L}_{\mathcal G}}\) to minimize the speaker similarity between the original and adversarial speech, where the subscript \(_{\mathcal{G}}\) denotes the generator. Meanwhile, the perturbation remover is guided by the loss function \(\mathcal{L}_{\mathcal{R}}\), which is defined between the adversarial perturbation \(\delta\) added to the original speech and its reverse form \(\delta'\). Here, the subscript \(_{\mathcal{R}}\) represents the remover. Notably, \({\mathcal L}_{\mathcal{G}}\) and  \({\mathcal L}_{\mathcal{R}}\) are defined in specific implementations. Finally, the overall loss function in the joint training framework is derived from the weighted summation of the speaker-adversarial and the restoration loss functions as follows:

\begin{equation}
\label{eq: joint training loss}
    {\mathcal{L}}=\beta {\mathcal{L}_{\mathcal G}} + \left(1-\beta\right){\mathcal{L}_{\mathcal R}},
\end{equation}
with $0\le\beta\le1$ to be the weight variable.  In this study, under the assumption that the remover \({\mathcal F}_{\mathcal R}\) is jointly trained with the generator \({\mathcal F}_{\mathcal G}\) an ideal configuration is investigated wherein \({\mathcal F}_{\mathcal G}\) and \({\mathcal F}_{\mathcal R}\) utilize the same SSED architecture.

\subsection{SSED-based generator\& remover}
\label{sec:SSED-J}
\begin{figure}[t]
    \centering
    \includegraphics[scale=0.55]{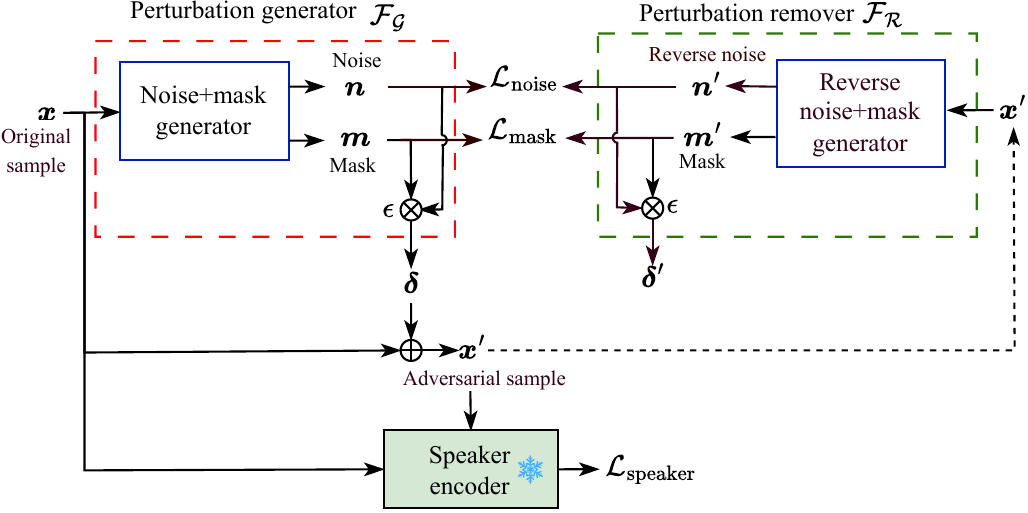}
    \caption{Joint training framework for the SSED-based speaker adversarial perturbation generator and remover. The Noise \& mask generator is inherited from the SSED architecture as presented in Fig. \ref{fig. SSED}. The speaker extractor is pre-trained and frozen.}
    \label{fig: SSED-based joint training}
    \end{figure}

As investigated in \cite{guo2024removal}, the architectures of the perturbation generator and remover are inherited from the SSED network as illustrated in Fig. \ref{fig: SSED-based joint training}. Given a sample from the original speech \(\boldsymbol{x}\), the noise \(\boldsymbol{n}\) and mask \(\boldsymbol{m}\) vectors are generated using the noise+mask generator. The adversarial perturbation \(\boldsymbol{\delta}\) is computed as the dot product between them as follows:
\begin{equation}
\boldsymbol{\delta}=\epsilon \cdot \left(\boldsymbol{n} \odot \boldsymbol{m}\right),    
\end{equation}
with $\epsilon$ to be the intensity factor. The adversarial sample \(\boldsymbol{x}'\) is generated with:
\begin{equation}
    \boldsymbol{x}' = \boldsymbol{x}+\boldsymbol{\delta}.
\end{equation}
Given an original utterance \(\mathcal{O}\), adversarial versions of all samples within it are generated accordingly, giving the adversarial utterance \(\mathcal{O}'\).

Following that, the speaker embedding vectors are extracted from \(\mathcal{O}\) and \(\mathcal{O}'\) using a pre-trained speaker extractor and represented with $\boldsymbol{v}$ and $\boldsymbol{v}'$, respectively. A speaker adversarial loss function is computed as the cosine similarity between $\boldsymbol{v}$ and $\boldsymbol{v}'$ as follows:
    \begin{equation}
    \label{eq:l_speaker}
    \mathcal{L}_{\rm speaker} = \frac{{\boldsymbol{v}}^{\mathsf T}\boldsymbol{v}'}{\lVert \boldsymbol{v} \lVert_{\rm 2} \lVert \boldsymbol{v}'\lVert_{\rm 2}},
    \end{equation}
with $\lVert \bullet \lVert_{\rm 2}$ denoting the L2 norm. By minimizing \({\mathcal L}_{\rm speaker}\), the speaker similarity between the adversarial and original utterances gets reduced.

Besides, a perceptual loss is defined on the generator accounting for preserving the quality of the adversarial speech. It is calculated on the adversarial sample ${\boldsymbol{x}'}$ and the mask vector ${\boldsymbol{m}}$ as follows:
\begin{equation}
    {\mathcal L}_{\rm perceptual} = \gamma \lVert \boldsymbol{x}'-\boldsymbol{x} \lVert_{\rm 2} + \left(1-\gamma\right) \lVert {\boldsymbol{m}} \lVert_{\rm 2},
    \label{eq. perceptual loss}
\end{equation}
where $0\le \gamma\le 1$ is the weight variable. The loss function for adversarial perturbation generation ${\mathcal L}_{\mathcal G}$ is derived as a balance between \(\mathcal{L}_{\rm{speaker}}\) and \(\mathcal{L}_{\rm{perceptual}}\) as follows:
\begin{equation}
    {\mathcal L}_{\mathcal G}=\eta\mathcal{L}_{\rm speaker}+\left(1-\eta\right){\mathcal L}_{\rm perceptual},
\label{eq: SSED generator}
\end{equation}
with $0\le\eta\le1$ serving as the weight variable.

Within the perturbation remover, given the adversarial sample \(\boldsymbol{x}'\), the reverse version of \(\boldsymbol{n}\) and \(\boldsymbol{m}\) are generated as \(\boldsymbol{n}'\) and \(\boldsymbol{m}'\), respectively. The restored version of \(\boldsymbol{x}\) is obtained as follows:
\begin{equation}
        \label{eq: restored sample}
        \hat{{\boldsymbol{x}}} = \boldsymbol{x}' + \underbrace{\epsilon\cdot( \boldsymbol{n}' \odot \boldsymbol{m}')}_{\boldsymbol{\delta}'},
\end{equation}
where ${\boldsymbol{\delta}}'$ is the reverse of ${\boldsymbol{\delta}}$.
As in the SSED network, the computation of \(\boldsymbol{\delta}\) is decomposed into the noise and mask vectors \(\boldsymbol{n}\) and \(\boldsymbol{m}\), the restoration loss function \(\mathcal{L}_{\text{restored}}\) is formulated with loss functions defined for each of them. In detail, ${\boldsymbol{n}}'$ is expected to be the reverse to ${\boldsymbol{n}}$, with the loss function defined between them as follows:
\begin{equation}
        {\mathcal L}_{\rm noise} = \lVert \boldsymbol{n} + {\boldsymbol{n}}'\lVert_{\rm 2}.
        \label{eq: noise}
\end{equation}
Meanwhile, ${\boldsymbol{m}}'$ is expected to be the same with ${\boldsymbol{m}}$, supervised by the loss function as follows:
\begin{equation}
         {\mathcal L}_{\rm mask} = \lVert \boldsymbol{m} - {\boldsymbol{m}}' \lVert_{\rm 2}.
         \label{eq: norm mask loss}
\end{equation}
The loss function accountable for the perturbation removal is computed as a combination of (\ref{eq: noise}) and (\ref{eq: norm mask loss}) as follows:
\begin{equation}
            {\mathcal L}_{\mathcal R} = \omega {\mathcal L}_{\rm mask} + \left(1-\omega\right) {\mathcal L}_{\rm noise},
\label{eq: Lrpt SSED}
\end{equation}
where the weight variable $0\le\omega\le 1$ balances between ${\mathcal L}_{\rm mask}$ and ${\mathcal L}_{\rm noise}$.

Finally, the loss function for the joint training framework utilizing the SSED architecture is derived by incorporating (\ref{eq: SSED generator}) and (\ref{eq: Lrpt SSED}) into (\ref{eq: joint training loss}).

\section{Experiments}
Our experiments were conducted to examine the removability of speaker-adversarial perturbations across the \textbf{ignorant}, \textbf{semi-informed}, and \textbf{well-informed} scenarios. Perturbations generated through both optimization-based and feedforward methods were assessed. The applicability of evaluation scenarios regarding speaker-adversarial perturbation generation and removal approaches is summarized in Table \ref{tab: Human perception and machine protection}.

\begin{table}

    \caption{Applicability of perturbation generation and removal algorithms regarding evaluation scenarios.}
    \label{tab: Human perception and machine protection}
    \centering

    \begin{subtable}[t]{\linewidth}
    \caption{Applicability of perturbation generation algorithms in the evaluation scenarios.}
    \label{tb: generation vs scenarios}
    \centering
    \begin{tabular}{cccc}
    \Xhline{1px}
    & Ignorant & Semi-informed & Well-informed \\
    \hline
    Optimization-based & \usym{2713} & \usym{2713} & \usym{2717} \\
    Feedforward & \usym{2713} & \usym{2713} & \usym{2713} \\
    \bottomrule
    \vspace{0.2cm}
    \end{tabular}
    \end{subtable}
    
    \begin{subtable}[t]{\linewidth}
    \centering
    \caption{Applicability of perturbation removal algorithms in the evaluation scenarios.}
    \label{tb: removal vs scenarios}

    \begin{tabular}{cccc}
    \Xhline{1px}
     & Ignorant & Semi-informed & Well-informed \\
    \hline
    Non-informed defense & \usym{2713} & \usym{2717} & \usym{2717} \\
    Denoising & \usym{2713} & \usym{2713} & \usym{2717} \\
    Joint training & - & - & \usym{2713} \\
    \Xhline{1px}
    \end{tabular}
    \end{subtable}
\end{table}

Table \ref{tab: Human perception and machine protection}(\subref{tb: generation vs scenarios}) illustrates the applicability of speaker-adversarial speech generation methods, including optimization-based and feedforward approaches, across the three evaluation scenarios. Between them, due to the lack of model training in optimization-based methods, they are not applicable in the well-informed scenario. Table \ref{tab: Human perception and machine protection}(\subref{tb: removal vs scenarios}) illustrates the applicability of the perturbation removal methodologies across the evaluation scenarios, including non-informed defense, denoising, and joint training methods. As the non-informed defense methods were developed without knowledge of the adversarial perturbation, they were only evaluated in the ignorant scenario. The denoising method was applied in both the ignorant and semi-informed scenarios. In the ignorant scenario, denoising models, unaware of speaker-adversarial perturbations, were trained using pairs of original and noisy utterances with additive noise. In the semi-informed scenario, where denoising models were assumed to be aware of speaker-adversarial speech, they were trained using pairs of original and adversarial utterances. The perturbation remover trained in the joint training framework was examined in the well-informed scenario.


\subsection{Datasets}
In our experiments, the VoxCeleb 1 \cite{nagrani2017voxceleb} and VoxCeleb 2 \cite{chung2018voxceleb2} datasets were employed to train speaker extractors, which were utilized for generating speaker-adversarial speech. Our experiments on speaker-adversarial speech were conducted on the LibriSpeech dataset. The recordings were resampled to 16 kHz. Specifically, the train-clean-100, train-clean-360, and train-other-500 partitions, providing 960 hours of utterances from 2,338 speakers in total, were used for training the speaker-adversarial perturbation generators and removers. The test-clean dataset was used for evaluation.

\subsection{Speaker encoder}
In our experiments on speaker-adversarial perturbation generation and removal, the speaker extractor adopted the encoder from the ECAPA-TDNN \cite{desplanques2020ecapa} architecture. The ECAPA-TDNN model was trained on the Voxceleb 1 \& 2 datasets, using the open-source ASV-subtools toolkit\footnote{\url{https://github.com/Snowdar/asv-subtools}} \cite{tong2021asv}. The MUSAN corpus \cite{snyder2015musan} and the RIR datasets \cite{ko2017study} were applied for data augmentation. The acoustic feature was a 40-dimensional filterbank with a 10 ms frameshift. Three SE-Res2Blocks were used. The dimension of the speaker embedding vector was 192.

\subsection{Evaluation metrics}
In our experiments, the speaker-adversarial perturbation removal algorithms were evaluated in terms of original speech restoration, speaker attribute restoration, and utility. Particularly, the restoration of original speaker attributes is defined as the recovery of their extraction, as examined in ASV evaluations. Utility was evaluated in speech quality, intelligibility, and prosody, respectively.

\subsubsection{Original speech restoration}
The restoration of original speech was measured using two metrics: scale-invariant signal-to-noise ratio (SI-SNR) and mean squared error (MSE). Both were calculated between the tested and original waveform samples. Higher SI-SNR and lower MSE indicate stronger restoration of the original speech. Specifically, in our experiments, an MSE value close to 0 and an SI-SNR value exceeding 50 were used as criteria of perturbation elimination and restoration of the original speech.

\subsubsection{ASV evaluations} ASV evaluations were conducted to measure the speaker attribute restoration capability in both white- and black-box conditions. In the white-box evaluations, the ECAPA-TDNN extractor used for speaker-adversarial speech generation was applied for speaker embedding extraction. In the black-box evaluation, the speaker encoder proposed in \cite{truong2024emphasized} was applied, referred to as \emph{ENSKD}. The open-source model \footnote{\url{https://github.com/ductuantruong/enskd}} was used for speaker embedding extraction. In both evaluations, cosine distance was computed as the score. Our ASV evaluations were carried out in a gender-dependent manner, with 20 male and 20 female speakers in the test-clean datasets, represented as \emph{libri-m} and \emph{libri-f}, respectively. For each speaker, one target and thirty nontarget trials were composed. The performances were measured with equal error rates (EERs).

In the ASV evaluations, the original utterance was used for speaker enrollment while the original, adversarial, and processed utterances were examined as test utterances, respectively. The EER obtained using the original utterance for testing was used as a reference. When testing with speaker-adversarial utterances, EERs higher than the reference EER indicated effective concealment of the original speaker's identity. When the processed utterance was utilized for testing, closer EERs to the reference EER indicated better restoration of speaker attributes.

\subsubsection{Speech quality} The speech quality was measured using the perceptual evaluation of speech quality (PESQ). The values were computed between the tested and the original speech utterances. The PESQ value ranged from 0 to 4.5. Higher PESQ values indicated less degradation in the speech quality of the tested utterances.

\subsubsection{Intelligibility}
The intelligibility was evaluated in the automatic speech recognition task (ASR). The Whisper\cite{radford2023robust} small English-only model \footnote{\url{https://github.com/openai/whisper}} provided by OpenAI was called for speech recognition. The performances were measured with word error rates (WERs). Using the WER of the original utterance as a reference, closer WER values from processed utterances to the reference WER indicated better perturbation removability.

\subsubsection{Prosody}
The pitch was applied as the indicator of the prosody attribute. Given a tested utterance and its original version, the pitch sequences were extracted, respectively. The Pearson correlation was computed between the two pitch sequences. Thereafter, the mean and standard deviation (STD) were calculated on the correlations obtained across the test set, represented as \emph{Pitch-Mean} and \emph{Pitch-STD} in the following. A higher Pitch-Mean and lower Pitch-STD indicated better pitch preservation. The open-source pitch extractor provided in \footnote{\url{https://github.com/Voice-Privacy-Challenge/Voice-Privacy-Challenge-2022}} was utilized.

\subsection{Speaker-adversarial speech generation}

In our experiments, speaker-adversarial speech utterances generated using both optimization-based, including MI-FGSM and GRA, and the feedforward SSED algorithms were examined. The configurations are detailed as follows.

\subsubsection{Optimization-based methods}
In the optimization-based methods, the loss function was defined on the cosine distance speaker embedding vectors extracted from the original and adversarial utterances using the ECAPA-TDNN speaker encoder, represented with ${\boldsymbol{y}}$ and ${\tilde{\boldsymbol{y}}}$, respectively. Mathematically, it is obtained as:
\begin{equation}
    \label{eq. optimization-based loss function}
        L\left(\bullet\right)=-\frac{{\boldsymbol y}^{\mathsf{T}}\tilde{{\boldsymbol y}}}{{\left\| {\boldsymbol{y}} \right\|_2}{\left\| {\tilde{\boldsymbol{y}}} \right\|_2}},
\end{equation}
with $\left\|\bullet\right\|_2$ to be the L2 norm. In particular, MI-FGSM and GRA algorithms were examined in our experiments for speaker-adversarial speech generation, respectively.

\subsubsection{SSED}
The speaker-adversarial speech generator was trained within the joint framework, as depicted in Fig. \ref{fig: joint training framework}, together with the remover. The generator and remover shared the same structure. The encoder consists of 3 convolutional blocks and 6 residual blocks. In each convolutional block, 1D convolution, batch normalization, and ReLU were applied. The kernel sizes of all convolutional layers were set to 7, 3, and 3, respectively. The noise decoder $\mathcal{G}_{\mathcal N}$ consisted of 3 transposed convolutional blocks, with kernel sizes of 3, 3, and 7, respectively. The tanh activation function was adopted in each layer. The same architecture was applied in the mask decoder ${\mathcal{G}_{\mathcal M}}$ while sigmoid was used for activation in the output layer. In the joint training process, the parameters were set as $\beta=0.94$, $\gamma=0.99$, $\eta=0.993$, $\omega=0.2$,  and $\epsilon=0.05$.

\setcounter{footnote}{0}
\renewcommand{\thefootnote}{\fnsymbol{footnote}}
    \begin{table*}[t]
        \caption{Evaluation results in the ignorant scenario where the non-informed defense and denoising methods are examined. The speaker-adversarial perturbations generated with the optimization-based MI-FGSM algorithm, and the feedforward SSED generator are included. Results obtained on original recording (\emph{Ori}) and speaker-adversarial (\emph{Adv}) speech by each method are provided for reference.\textsuperscript{*}}
        
        \label{tb. ignorant}
        \centering
        \scriptsize

        \begin{subtable}[t]{\linewidth}
        \centering
        \caption{Original speech restoration evaluation results. SI-SNR and MSE values are included. \emph{N.A.} is short for \emph{not applicable}.}
        \label{tb. ignorant restoration res}
        \begin{tabular}{c|c|c|cccccc|c|cccccc}
        \Xhline{1px}
        \multirow{4}{*}{}&\multirow{3}{*}{Ori}&\multicolumn{7}{c|}{MI-FGSM}&\multicolumn{7}{c}{SSED}\\
         \cmidrule(r){3-9} \cmidrule(r){10-16}
        &&\multirow{2}{*}{Adv}&\multicolumn{5}{c}{Non-informed Defense}& Denoising &\multirow{2}{*}{Adv}&\multicolumn{5}{c}{Non-informed defense}& Denoising\\
        \cmidrule(r){4-8}\cmidrule(r){9-9}\cmidrule(r){11-15}\cmidrule(r){16-16}
        &&&QT & AAC & MS &AN &GAN&FRCRN-N&&QT &AAC& MS& AN &GAN&FRCRN-N\\
        \Xhline{1px}
        
        SI-SNR&$\cellcolor{black!0}\infty$& \cellcolor{black!24}31.23&\cellcolor{black!36}26.73 &N.A.& \cellcolor{black!60}15.41&\cellcolor{black!48}23.92&N.A.&\cellcolor{black!18}31.29&\cellcolor{black!6}32.43&\cellcolor{black!30}26.95&N.A.&\cellcolor{black!54}15.44&\cellcolor{black!42}24.14&N.A.&\cellcolor{black!12}32.38\\
        MSE&\cellcolor{black!0}0.00&\cellcolor{black!24}22.67&\cellcolor{black!36}64.43&N.A.&\cellcolor{black!54}1388.98&\cellcolor{black!48}138.93&N.A.&\cellcolor{black!18}22.38&\cellcolor{black!6}17.45&\cellcolor{black!30}61.08&N.A.&\cellcolor{black!60}1451.46&\cellcolor{black!42}133.70&N.A.&\cellcolor{black!12}17.70\\
        
        \Xhline{1px}
        \end{tabular}
        \vspace{0.3cm}
        \end{subtable}


        \begin{subtable}[t]{\linewidth}
        \centering
        \caption{EERs(\%) obtained in white- and black-box ASV evaluations. Results are presented for male and female trials, denoted as libri-m and libri-f, respectively, where \emph{libri} refers to the LibriSpeech dataset.}
        \begin{tabular}{c | c|c | cccccc | c | cccccc}
        \Xhline{1px}
        \multirow{4}{*}{Trials}&\multirow{3}{*}{Ori}&\multicolumn{7}{c|}{MI-FGSM}&\multicolumn{7}{c}{SSED}\\
         \cmidrule(r){3-9} \cmidrule(r){10-16}
        &&\multirow{2}{*}{Adv}&\multicolumn{5}{c}{Non-informed Defense}& Denoising &\multirow{2}{*}{Adv}&\multicolumn{5}{c}{Non-informed Defense}& Denoising\\
        \cmidrule(r){4-8}\cmidrule(r){9-9}\cmidrule(r){11-15}\cmidrule(r){16-16}
        &&&QT &AAC& MS& AN &GAN&FRCRN-N&&QT &AAC& MS& AN &GAN&FRCRN-N\\
        \Xhline{1px}
        \multicolumn{16}{c}{White-box}\\ 
        \hline
        libri-m&\cellcolor{black!0}1.21&\cellcolor{black!60}56.74&\cellcolor{black!45}18.51&\cellcolor{black!52}21.58&\cellcolor{black!55}38.66&\cellcolor{black!32}13.03&\cellcolor{black!37}13.69&\cellcolor{black!58}56.63&\cellcolor{black!46}18.62&\cellcolor{black!4}3.94&\cellcolor{black!31}12.60&\cellcolor{black!42}18.07&\cellcolor{black!6}4.92&\cellcolor{black!17}8.65&\cellcolor{black!42}18.07\\
        libri-f&\cellcolor{black!1}1.64&\cellcolor{black!57}53.72&\cellcolor{black!44}18.28&\cellcolor{black!53}22.03&\cellcolor{black!54}37.26&\cellcolor{black!40}15.17&\cellcolor{black!38}13.71&\cellcolor{black!56}53.13&\cellcolor{black!51}19.27&\cellcolor{black!8}5.27&\cellcolor{black!36}13.59&\cellcolor{black!50}18.92&\cellcolor{black!5}4.80&\cellcolor{black!15}7.26&\cellcolor{black!50}18.92\\ 
        \Xhline{1px}
        \multicolumn{16}{c}{Back-box}\\ 
        \hline
        libri-m&\cellcolor{black!0}1.21&\cellcolor{black!41}18.01&\cellcolor{black!19}9.09&\cellcolor{black!16}8.43&\cellcolor{black!34}13.36&\cellcolor{black!10}5.92&\cellcolor{black!39}13.93&\cellcolor{black!43}18.10&\cellcolor{black!30}12.17&\cellcolor{black!12}6.65&\cellcolor{black!25}10.73&\cellcolor{black!7}4.93&\cellcolor{black!3}3.65&\cellcolor{black!21}10.17&\cellcolor{black!27}11.72\\
        libri-f&\cellcolor{black!2}2.23&\cellcolor{black!48}18.86&\cellcolor{black!24}10.55&\cellcolor{black!22}10.21&\cellcolor{black!33}13.24&\cellcolor{black!14}7.21&\cellcolor{black!36}13.59&\cellcolor{black!49}18.89&\cellcolor{black!28}11.78&\cellcolor{black!18}8.96&\cellcolor{black!26}11.63&\cellcolor{black!13}6.85&\cellcolor{black!9}5.39&\cellcolor{black!20}9.35&\cellcolor{black!29}11.83\\  
        \Xhline{1px}
         \end{tabular}
        \label{tb. ignorant ASV res}
        \vspace{0.3cm}
        \end{subtable}


        \begin{subtable}[t]{\linewidth}
        \centering
        \caption{Utility evaluation results. PESQ, WER, Pitch-Mean, and Pitch-STD values are included.}
        \label{tb. ignorant utility res}
        \begin{tabular}{c | c|c | cccccc | c | cccccc}
        \Xhline{1px}
        \multirow{4}{*}{}&\multirow{3}{*}{Ori}&\multicolumn{7}{c|}{MI-FGSM}&\multicolumn{7}{c}{SSED}\\
         \cmidrule(r){3-9} \cmidrule(r){10-16}
        &&\multirow{2}{*}{Adv}&\multicolumn{5}{c}{Non-informed Defense}& Denoising &\multirow{2}{*}{Adv}&\multicolumn{5}{c}{Non-informed Defense}& Denoising\\
        \cmidrule(r){4-8}\cmidrule(r){9-9}\cmidrule(r){11-15}\cmidrule(r){16-16}
        &&&QT &AAC& MS& AN &GAN&FRCRN-N&&QT &AAC& MS& AN &GAN&FRCRN-N\\
        \Xhline{1px}
        PESQ&\cellcolor{black!0}4.50&\cellcolor{black!15}3.64&\cellcolor{black!20}3.31&\cellcolor{black!50}3.08&\cellcolor{black!25}3.28&\cellcolor{black!35}3.16&\cellcolor{black!55}2.20&\cellcolor{black!10}3.65&\cellcolor{black!5}3.67&\cellcolor{black!30}3.26&\cellcolor{black!45}3.09&\cellcolor{black!20}3.31&\cellcolor{black!40}3.15&\cellcolor{black!60}2.19&\cellcolor{black!5}3.67\\
        WER&\cellcolor{black!0}4.08&\cellcolor{black!12}4.82&\cellcolor{black!8}4.75&\cellcolor{black!38}5.53&\cellcolor{black!30}5.21&\cellcolor{black!4}4.68&\cellcolor{black!55}8.75&\cellcolor{black!17}4.87&\cellcolor{black!51}6.94&\cellcolor{black!25}5.17&\cellcolor{black!47}6.49&\cellcolor{black!42}5.61&\cellcolor{black!21}5.01&\cellcolor{black!60}9.47&\cellcolor{black!34}5.26\\
        Pitch-Mean&\cellcolor{black!0}1.00&\cellcolor{black!8}0.99&\cellcolor{black!34}0.95&\cellcolor{black!51}0.90&\cellcolor{black!25}0.97&\cellcolor{black!42}0.94&\cellcolor{black!60}0.65&\cellcolor{black!8}0.99&\cellcolor{black!8}0.99&\cellcolor{black!17}0.98&\cellcolor{black!51}0.90&\cellcolor{black!34}0.95&\cellcolor{black!34}0.95&\cellcolor{black!60}0.65&\cellcolor{black!8}0.99\\
        Pitch-STD&\cellcolor{black!0}0.00&\cellcolor{black!13}0.04&\cellcolor{black!20}0.06&\cellcolor{black!33}0.09&\cellcolor{black!26}0.08&\cellcolor{black!46}0.12&\cellcolor{black!53}0.19&\cellcolor{black!13}0.04&\cellcolor{black!6}0.03&\cellcolor{black!60}0.43&\cellcolor{black!26}0.08&\cellcolor{black!26}0.08&\cellcolor{black!40}0.11&\cellcolor{black!53}0.19&\cellcolor{black!6}0.03\\
        \Xhline{1px}
        \end{tabular}
        \end{subtable}
        
    \end{table*}

\setcounter{footnote}{4}
\renewcommand{\thefootnote}{\arabic{footnote}}

\subsection{Ignorant methods configurations}
The non-informed defense and denoising methods were evaluated in the ignorant scenario, detailed as follows:

\subsubsection{Non-informed defense methods}
In our experiments on non-informed defense methods, quantization (QT), AAC compression, median smoothing (MS), adding noise (AN), and a GAN-based generative model were examined. The specific parameters for each method were configured as follows:

\expandafter{\romannumeral 1}. \emph{QT}\cite{characterinzing}: In the QT method, a quantization level $\lambda$ of $2^8$ was applied.

\expandafter{\romannumeral 2}. \emph{AAC}\cite{chen2022towards}: The AAC compression was applied.

\expandafter{\romannumeral 3}. \emph{MS}\cite{characterinzing}: The MS method adopted the kernel size of 3.

\expandafter{\romannumeral 4}. \emph{AN}\cite{a15080293}: The Gaussian noise of SNR=25dB was utilized.


\expandafter{\romannumeral 5}. \emph{GAN}\cite{9551961}: The Parallel WaveGAN Defense model described in \cite{9551961} was applied. The Parallel WaveGAN \cite{9053795} consisted of 30 layers of dilated
 residual convolution blocks with exponentially increasing three dilation cycles \cite{vandenoord16_ssw}. The number of residual and skip channels were set to 64 and the convolution filter size was set to three. The loss function for the generator is a linear combination of the multi-resolution STFT loss and the adversarial loss \cite{9053795}.

The QT, AAC, MS, and AN methods were implemented using the open-source toolkit\footnote{\url{https://github.com/speakerguard/speakerguard}}, while the GAN model was trained with the open-source toolkit\footnote{\url{https://github.com/kan-bayashi/ParallelWaveGAN}}.

\subsubsection{Denoising methods}
The denoising model applied in the ignorant scenario adopted the FRCRN architecture\cite{9747578}.  The model was trained with the open-source toolkit \footnote{\url{https://modelscope.cn/models/damo/speech_frcrn_ans_cirm_16k/summary}}, using pairs of clean and noisy speech utterances. Noisy utterances were obtained by adding noise to clean utterances, with a signal-to-noise ratio ranging from 28 to 36 dB. The NOISEX-92 corpus \cite{NOISEX-92} was utilized for the noise set. The model is referred to as \emph{FRCRN-N}, with \emph{N} indicating \emph{noise}.

\setcounter{footnote}{0}
\renewcommand{\thefootnote}{\fnsymbol{footnote}}

\begin{table*}[t]
    
    \caption{Evaluation results in the semi-informed scenario. The DCCRN-A, FRCRN-A, and Denoising-G models are included. The adversarial speech generation methods of MI-FGSM, GRA, and SSED are examined. Results obtained on original recording (\emph{Ori}) and speaker-adversarial (\emph{Adv}) speech by each method are provided for reference.\textsuperscript{*}}
    \label{tb. semi-informed}
    \centering
    \scriptsize



    \begin{subtable}[t]{\linewidth}
    \centering
    \caption{Original speech restoration evaluation results. SI-SNR and MSE values are included.}
    \begin{tabular}{c | c | c | cc | c | cc | c|ccc }
    \Xhline{1px}
    \multirow{2}{*}{}&\multirow{2}{*}{Ori}& \multicolumn{3}{c|}{MI-FGSM}& \multicolumn{3}{c|}{GRA}& \multicolumn{4}{c}{SSED}\\
    \cmidrule(r){3-5} \cmidrule(r){6-8} \cmidrule(r){9-12}
    &&Adv&DCCRN-A&FRCRN-A&Adv&DCCRN-A&FRCRN-A&Adv&DCCRN-A&FRCRN-A &Denoising-G\\    
    \Xhline{1px}
    SI-SNR&$\cellcolor{black!0}\infty$&\cellcolor{black!60}31.23&\cellcolor{black!54}31.45&\cellcolor{black!30}35.10&\cellcolor{black!36}33.25&\cellcolor{black!42}33.22&\cellcolor{black!24}36.44&\cellcolor{black!48}32.43&\cellcolor{black!12}40.51&\cellcolor{black!18}40.02&\cellcolor{black!6}42.64\\
    MSE&\cellcolor{black!0}0.00&\cellcolor{black!48}22.67&\cellcolor{black!60}74.38&\cellcolor{black!30}9.78&\cellcolor{black!36}14.23&\cellcolor{black!54}70.43&\cellcolor{black!24}7.15&\cellcolor{black!42}17.45&\cellcolor{black!12}2.77&\cellcolor{black!18}3.16&\cellcolor{black!6}1.80\\

    \Xhline{1px}
    \end{tabular}
    \vspace{0.3cm}
    \end{subtable}


    \begin{subtable}[t]{\linewidth}
    \centering
    \caption{EERs(\%) obtained in white- and black-box ASV evaluations. Results are presented on libri-m and libri-f, respectively.}
    \begin{tabular}{c | c | c | cc | c | cc | c|ccc }
    \Xhline{1px}
    \multirow{2}{*}{Trials}&\multirow{2}{*}{Ori}& \multicolumn{3}{c|}{MI-FGSM}& \multicolumn{3}{c|}{GRA}& \multicolumn{4}{c}{SSED}\\
    \cmidrule(r){3-5} \cmidrule(r){6-8} \cmidrule(r){9-12}
    &&Adv&DCCRN-A&FRCRN-A&Adv&DCCRN-A&FRCRN-A&Adv&DCCRN-A&FRCRN-A &Denoising-G\\
    \Xhline{1px}
    \multicolumn{12}{c}{White-box}\\ 
    \hline
    libri-m&\cellcolor{black!0}1.21&\cellcolor{black!60}56.74&\cellcolor{black!53}32.42&\cellcolor{black!49}30.67&\cellcolor{black!52}32.09&\cellcolor{black!33}10.73&\cellcolor{black!40}13.58&\cellcolor{black!44}18.62&\cellcolor{black!1}1.31&\cellcolor{black!1}1.31&\cellcolor{black!0}1.21\\
    libri-f&\cellcolor{black!6}1.64&\cellcolor{black!58}53.72&\cellcolor{black!56}33.04&\cellcolor{black!50}31.34&\cellcolor{black!55}32.57&\cellcolor{black!36}11.89&\cellcolor{black!41}16.05&\cellcolor{black!47}19.27&\cellcolor{black!10}2.05&\cellcolor{black!7}1.82&\cellcolor{black!9}1.99\\
    \Xhline{1px}
    \multicolumn{12}{c}{Black-box}\\ 
    \hline
    libri-m&\cellcolor{black!0}1.21&\cellcolor{black!43}18.01&\cellcolor{black!26}6.24&\cellcolor{black!21}5.50&\cellcolor{black!29}8.43&\cellcolor{black!18}3.23&\cellcolor{black!20}3.94&\cellcolor{black!38}12.17&\cellcolor{black!1}1.31&\cellcolor{black!3}1.36&\cellcolor{black!4}1.37\\
    libri-f&\cellcolor{black!12}2.23&\cellcolor{black!46}18.86&\cellcolor{black!30}9.49&\cellcolor{black!27}7.56&\cellcolor{black!32}10.01&\cellcolor{black!23}5.92&\cellcolor{black!24}6.21&\cellcolor{black!35}11.78&\cellcolor{black!16}2.64&\cellcolor{black!13}2.40&\cellcolor{black!15}2.52\\
    \Xhline{1px}
    \end{tabular}
    \vspace{0.3cm}
    \end{subtable}


    \begin{subtable}[t]{\linewidth}
    \centering
    \caption{Utility evaluation results. PESQ, WER, Pitch-Mean, and Pitch-STD values are included.}
    \begin{tabular}{c | c | c | cc | c | cc | c|ccc }
    \Xhline{1px}
    \multirow{2}{*}{}&\multirow{2}{*}{Ori}& \multicolumn{3}{c|}{MI-FGSM}& \multicolumn{3}{c|}{GRA}& \multicolumn{4}{c}{SSED}\\
    \cmidrule(r){3-5} \cmidrule(r){6-8} \cmidrule(r){9-12}
    &&Adv&DCCRN-A&FRCRN-A&Adv&DCCRN-A&FRCRN-A&Adv&DCCRN-A&FRCRN-A &Denoising-G\\    
    \Xhline{1px}
    PESQ&\cellcolor{black!0}4.50 & \cellcolor{black!60}3.64 &\cellcolor{black!36}4.00&\cellcolor{black!30}4.03&\cellcolor{black!48}3.81&\cellcolor{black!42}3.88&\cellcolor{black!24}4.11&\cellcolor{black!54}3.67&\cellcolor{black!18}4.19&\cellcolor{black!12}4.31&\cellcolor{black!6}4.36\\
    WER&\cellcolor{black!0}4.08&\cellcolor{black!54}4.82&\cellcolor{black!42}4.52&\cellcolor{black!24}4.45&\cellcolor{black!12}4.21&\cellcolor{black!48}4.78&\cellcolor{black!18}4.40&\cellcolor{black!60}6.94&\cellcolor{black!36}4.49&\cellcolor{black!30}4.46&\cellcolor{black!6}4.12\\
    Pitch-Mean&\cellcolor{black!0}1.00&\cellcolor{black!15}0.99&\cellcolor{black!60}0.94&\cellcolor{black!45}0.95&\cellcolor{black!15}0.99&\cellcolor{black!60}0.94&\cellcolor{black!30}0.96&\cellcolor{black!15}0.99&\cellcolor{black!30}0.96&\cellcolor{black!30}0.96&\cellcolor{black!0}1.00\\
    Pitch-STD&\cellcolor{black!0}0.00&\cellcolor{black!36}0.04&\cellcolor{black!60}0.07&\cellcolor{black!48}0.05&\cellcolor{black!24}0.03&\cellcolor{black!60}0.07&\cellcolor{black!48}0.05&\cellcolor{black!24}0.03&\cellcolor{black!48}0.05&\cellcolor{black!48}0.05&\cellcolor{black!12}0.01\\
    \Xhline{1px}
    \end{tabular}
    \end{subtable}

\end{table*}

\setcounter{footnote}{7}
\renewcommand{\thefootnote}{\arabic{footnote}}

\subsubsection{Results}
\label{sec: Removal MI-FGSM}
Experimental results for the ignorant scenario, using speaker-adversarial speech generated by the MI-FGSM, and SSED algorithms, are presented in Table \ref{tb. ignorant}. In detail, Fig. \ref{tb. ignorant}(\subref{tb. ignorant restoration res}) presents the results of the restoration of the original speech. The EERs obtained in ASV evaluations are given in Table \ref{tb. ignorant}(\subref{tb. ignorant ASV res}), measuring the speaker attribute restoration capability. The utility results are given in Table \ref{tb. ignorant}(\subref{tb. ignorant utility res}).


First, as shown in Table \ref{tb. ignorant}(\subref{tb. ignorant restoration res}), adversarial speech generated using the MI-FGSM and SSED methods demonstrated comparable intensity, reflected in their SI-SNR and MSE values. This established a basis for a fair comparison between these two methods. In the ASV evaluation results, as provided in Table \ref{tb. ignorant}(\subref{tb. ignorant ASV res}), adversarial utterances generated by the MI-FGSM and SSED methods showed higher EERs compared to original speech in both male and female trials, demonstrating the effectiveness of adversarial perturbations in distorting speaker attributes. Besides, as presented in Table \ref{tb. ignorant}(\subref{tb. ignorant utility res}), compared with the original speech, the PESQ values decreased for the adversarial speech generated by both methods, indicating the adverse effects of the perturbations on speech quality. In the ASR evaluations, the adversarial speech exhibited a higher WER compared to the original version, suggesting the effects of the perturbations on speech intelligibility. In the prosody evaluation, the adversarial speech exhibited marginally lower Pitch-Mean and higher Pitch-STD values, indicating a minor influence of the perturbations on the prosodic characteristic.

Compared with the adversarial speech utterances, the experimental results in Table \ref{tb. ignorant} demonstrate the efficacy of non-informed defense methods and the denoising model in perturbation removal as follows:

\renewcommand{\thefootnote}{\fnsymbol{footnote}}
\footnotetext[1]{In the SI-SNR, MSE, PESQ, WER, Pitch-Mean, and Pitch-STD metrics, the background color of each cell indicates the relative performance within each row. For the EER values, the background color is set within the white-box and black-box scenarios, respectively. Lighter colors indicate values closer to those obtained on the original recording.}
\renewcommand{\thefootnote}{\arabic{footnote}}

Among non-informed defense methods, the AAC method resulted in information loss, while the GAN method produced a new speech utterance that could not be aligned with the original in waveform samples. Consequently, the SI-SNR and MSE metrics were not applied for both methods. When applying the remaining non-informed methods to adversarial speech generated with MI-FGSM, the processed utterances exhibited lower SI-SNR values than the adversarial counterparts. Moreover, the evaluated non-informed defense methods yielded higher MSE values than the adversarial utterances. Such observations indicate that the evaluated non-informed defense methods were negative to the restoration of the original speech. In terms of the impacts on speaker attribute extraction, the processed speech yielded lower EERs than adversarial speech in ASV evaluations conducted in both white- and black-box scenarios. This demonstrates their effectiveness in alleviating the adversarial perturbations on speaker attributes. However, the EERs were still higher than those obtained on the original utterance, indicating that these methods did not restore the original speaker attributes present in the utterances. Moreover, regarding speech quality, lower PESQ values than the adversarial speech were obtained, indicating their adverse effects on the speech quality. In terms of intelligibility, the QT and AN methods achieved slightly lower WERs than the adversarial speech but still higher than the original version, implying their incapability of eliminating the effects of the perturbations on the linguistic content. The other defense methods obtained higher WERs than the adversarial speech, indicating negative effects on the intelligibility of the restored speech. In terms of pitch extraction, the non-informed defense methods achieved lower Pitch-Mean and higher Pitch-STD compared to the adversarial speech, implying their adverse impact on pitch extraction.

On the SSED-generated speaker-adversarial speech, consistent results were observed in the non-informed defense methods with MI-FGSM across all evaluation metrics, except for WER. On the speaker-adversarial speech generated by the SSED generator, WERs increased across all non-informed defense methods, indicating their adverse effects on the linguistic content within the speech utterances. In summary, the examined non-informed defense methods were able to mitigate the effects of speaker-adversarial perturbations in speaker attribute extraction. However, they were unable to restore original speech, as evidenced by the MSE values much higher than 0, higher EER, WER, and Pitch-STD values, along with reduced PESQ, SI-SNR, and Pitch-Mean values compared with the original speech.

Moreover, on the speaker-adversarial speech generated with the MI-FGSM method, using the FRCRN-N denoising model, the processed speech achieved comparable SI-SNR, MSE, EER, PESQ, and WER values with the adversarial versions. These results indicate that the FRCRN-N model was unable to reduce speaker-adversarial perturbations from the adversarial speech utterances, nor mitigate their effects on speaker attribute extraction and utility. Similar results were observed on the speaker-adversarial speech generated by the SSED generator, further demonstrating that the denoising model trained on additive noise failed to remove speaker-adversarial perturbations.

Overall, in the ignorant scenario, the effects of speaker-adversarial perturbations on speaker attribute extraction can be mitigated by non-informed defense methods, whereas the denoising model FRCRN-N was proved ineffective in this regard. Furthermore, both the non-informed defense methods and the FRCRN-N denoising model were unable to remove perturbations or enhance the utility of adversarial speech.

\subsection{Semi-informed evaluations}

The semi-informed experiments were conducted on the speaker-adversarial speech generated using optimization-based methods, including MI-FGSM and GRA, as well as the feedforward SSED generator. The denoising models were examined, including the DCCRN and FRCRN architectures, which were trained using pairs of original and adversarial speech utterances. These denoising models are represented with \emph{DCCRN-A} and \emph{FRCRN-A}, respectively, with \emph{A} short for \emph{adversarial}. Additionally, to investigate the potential for semi-informed perturbation removal on speaker-adversarial speech generated by the SSED generator, a denoising model with the architecture of the SSED generator was trained on paired original and adversarial speech samples. The denoising model is represented with \emph{denoising-G} with \emph{G} short for \emph{generator}. The detailed configurations of the denoising models are as follows:
\subsubsection{DCCRN-A}
A DCCRN-A model was trained on speaker-adversarial speech generated by the SSED, MI-FGSM, and GRA algorithms, respectively. Each model was evaluated on its corresponding adversarial utterances. The architecture of DCCRN-CL \cite{hu20g_interspeech} was applied for all the models. The channel sizes of convolutional layers in the encoder of the DCCRN-CL were \{32, 64, 128, 256, 256, 256\}. Besides, DCCRN-CL adopted complex
LSTM \cite{hu20g_interspeech} with 128 units for the real part and imaginary
part, respectively. Reference for the model architecture can be found at \footnote{\url{https://github.com/huyanxin/DeepComplexCRN}}.

\subsubsection{FRCRN-A}
An FRCRN-A model was trained on speaker-adversarial speech generated by the SSED, MI-FGSM, and GRA algorithms, respectively. Evaluations were conducted on their corresponding adversarial utterances. In detail, the number of CR blocks \cite{tan2018convolutional} in the encoder and decoder of CRED was 6. The time-recurrent module was composed of 2 stacked CFSMN \cite{tan2018convolutional} layers. Each CR block used 128 channels for convolution operations. 128 units were used for the real and imaginary parts of the CFSMN. Only the SI-SNR loss function was used in model training. Reference for the model architecture can be found at\footnote{\url{https://www.modelscope.cn/models/iic/speech_frcrn_ans_cirm_16k}}.

\subsubsection{Denoising-G}
A denoising model, adopting the architecture of the SSED generator, was trained and evaluated on SSED-generated speaker-adversarial speech. In the training process, the parameters were set as $\gamma=0.99$, $\eta=0.993$, $\omega=0.2$, and $\epsilon=0.05$.

\subsubsection{Results}
The semi-informed evaluation results, including original speech restoration, ASV performance, and utility metrics, are summarized in Table \ref{tb. semi-informed}. In ASV evaluations on the speaker-adversarial speech generated by the MI-FGSM and GRA algorithms, utterances processed by DCCRN-A and FRCRN-A models achieved lower EERs than adversarial speech. This demonstrates the effectiveness of the DCCRN-A and FRCRN-A models in mitigating the impact of the adversarial perturbations on speaker attribute extraction. However, complete removal of speaker-adversarial perturbations from the speech remains unrealized. This was evidenced by MSE values much higher than 0, SI-SNR values below 50, higher EERs compared to the original recordings in ASV evaluations, PESQ scores under 4.5, higher WERs, lower Pitch-Means, and higher Pitch-STDs than the original recordings.

On the speaker-adversarial speech generated by the SSED algorithm, the DCCRN-A and FRCRN-A models achieved better performances than on those generated with the optimization-based methods. This may be attributed to the fact that, unlike the iterative strategy employed in the optimization-based methods, the SSED utilized a feedforward architecture for adversarial speech generation. This facilitated the easier learning of the SSED-generated perturbations by the DCCRN-A and FRCRN-A models. Moreover, the denoising model utilizing the SSED generator architecture (Denoising-G) achieved lower MSE and higher SI-SNR values compared to the DCCRN-A and FRCRN-A models. This demonstrates its superior perturbation removal capability, suggesting that knowledge of the perturbation generator architecture is beneficial to its removal. This conclusion is further supported by its higher PESQ, lower WER, higher Pitch-Mean, and lower Pitch-STD values compared to the DCCRN-A and FRCRN-A models. In white-box ASV evaluations, the Denoising-G model achieved lower EERs than DCCRN-A and FRCRN-A, demonstrating superior performance in reducing the impact of speaker-adversarial perturbations on speaker attribute extraction.

Above all, the evaluation results obtained in the semi-informed scenario show that: 1) the denoising models, including DCCRN-A, FRCRN-A, and Denoising-G, were effective in mitigating the effects of speaker-adversarial perturbations on speaker attribute extraction and utility, 2) the semi-informed models were unable to eliminate the perturbations and restore the original speech, especially those generated with the optimization-based methods, 3) on the speaker-adversarial speech generated with the SSED model, as indicated by the results obtained by the Denoising-G model, knowledge of the architecture of the generator was beneficial to perturbation removal, 4) using the Denoising-G model, the speaker-adversarial perturbations generated by the feedforward SSED model can be largely removed thus considerably restoring the original speech.

\subsection{Well-informed evaluations}

\begin{table}[t]
    \caption{Evaluation results in the well-informed scenario. Results obtained on original recording (\emph{Ori}) and speaker-adversarial (\emph{Adv}) speech by each method are provided for reference. The processed speech is shown in the column of \emph{Proce}.}
    \label{tb. well-informed}
    \centering
    \scriptsize

    \begin{subtable}[t]{\linewidth}
    \centering
    \caption{Original speech restoration evaluation results. SI-SNR and MSE values are included.}
    \begin{tabular}{c |c | c | c  }
    \Xhline{1px}
   &Ori&Adv&Proce\\
    \Xhline{1px}
    SI-SNR&$\infty$&32.43&50.29\\
    MSE&0&1.80&0.28\\
    \Xhline{1px}
    \end{tabular}
    \vspace{0.3cm}
    \end{subtable}

    \begin{subtable}[t]{\linewidth}
    \centering
    \caption{EERs(\%) obtained in white- and black-box ASV evaluations. Results are presented on libri-m and libri-f, respectively.}
    \begin{tabular}{ c | c | c | c }
    \Xhline{1px}

   Trials &Ori&Adv&Proce\\
    \Xhline{1px}
     \multicolumn{4}{c}{White-box}\\ 
    \hline
    libri-m&1.21&18.62&1.21\\
    libri-f&1.64&19.27&1.64\\
    \hline
    \multicolumn{4}{c}{Black-box}\\ 
    \hline
    libri-m&1.21&12.17&1.15\\
    libri-f&2.23&11.78&2.29\\
    \Xhline{1px}
    \end{tabular}
    \vspace{0.3cm}
    \end{subtable}

    \begin{subtable}[t]{\linewidth}
    \centering
    \caption{Utility evaluation results. PESQ, WER, Pitch-Mean, and Pitch-STD values are included.}
    \begin{tabular}{c |c | c | c  }
    \Xhline{1px}
   &Ori&Adv&Proce\\
    \Xhline{1px}
    PESQ&4.50&3.67&4.47\\
    WER&4.08&6.94&4.08\\
    Pitch-Mean&1.00&0.99&1.00\\
    Pitch-STD&0&0.03&0.01\\
    \Xhline{1px}
    \end{tabular}
    \end{subtable}
\end{table}

The well-informed evaluations were performed on the remover that was jointly trained with the generator, utilizing the SSED architecture. The evaluation results are shown in Table \ref{tb. well-informed}. Firstly, an SI-SNR value of 50.29, exceeding 50, and an MSE value of 0.28, close to 0, indicated that the perturbations were nearly removed, and the original speech was almost restored. In terms of speaker attribute extraction, the processed speech achieved lower EERs than the adversarial speech, demonstrating its effectiveness in reducing the impacts of speaker-adversarial perturbations on speaker attribute extraction. Furthermore, compared to the original recording, the processed speech achieved equivalent EERs in white-box evaluations and comparable EERs in black-box evaluations. In the speech quality evaluations, the processed speech achieved a PESQ value of 4.47, nearing the optimal value of 4.5. In the intelligibility assessment, it achieved a WER of 4.08, equivalent to the original recording. In the prosody evaluation, the Pitch-Mean and Pitch-STD values were 1.00 and 0.01, respectively, indicating minimal impact on pitch. In summary, in the well-informed evaluations, where the perturbation generator and remover were jointly trained, the speaker-adversarial perturbations were nearly removed, thereby restoring the original recording.

\setcounter{footnote}{0}
\renewcommand{\thefootnote}{\fnsymbol{footnote}}

\subsection{Cross-domain evaluations}

    \begin{table*}[t]
    \caption{Evaluation results obtained on the cross-domain datasets, VoxCeleb and AIShell, in the ignorant, semi-informed and well-informed scenarios. The adversarial speech is generated with SSED. Results obtained on original recording (\emph{Ori}) and speaker-adversarial (\emph{Adv}) speech by each method are provided for reference.\textsuperscript{*}}
    \label{tb. cross domain}
    \scriptsize

    \begin{subtable}[h]{\linewidth}
    \caption{Original speech restoration evaluation results. SI-SNR and MSE values are included.}
        \centering
        \begin{tabular}{c | c | cc | ccccc | cc c | c}
        \Xhline{1px}
        \multirow{3}{*}{}&\multirow{3}{*}{}&\multirow{3}{*}{Ori}&\multirow{3}{*}{Adv} & \multicolumn{5}{c|}{Ignorant}&\multicolumn{3}{c|}{\multirow{2}{*}{Semi-informed}}&\multirow{3}{*}{Well-informed}\\
        \cmidrule(r){5-9}
        &&&&\multicolumn{4}{c}{Non-informed Defense}& Denoising &\multicolumn{3}{c|}{}&\\
        \cmidrule(r){5-8}\cmidrule(r){9-9}\cmidrule(r){10-11} \cmidrule(r){12-12}
        &&&&QT &AAC& MS& AN&FRCRN-N&DCCRN-A&FRCRN-A&Denoising-G&\\
        \Xhline{1px}
        \multirow{2}{*}{VoxCeleb}&SI-SNR&$\cellcolor{black!0}\infty$&\cellcolor{black!33}30.59&\cellcolor{black!46}24.34&$-$&\cellcolor{black!60}16.99&\cellcolor{black!53}23.20&\cellcolor{black!40}30.11&\cellcolor{black!26}34.76&\cellcolor{black!20}36.97&\cellcolor{black!13}41.16&\cellcolor{black!6}47.81\\
        &MSE&\cellcolor{black!0}0.00&\cellcolor{black!26}15.76&\cellcolor{black!46}63.21&$-$&\cellcolor{black!60}1383.41&\cellcolor{black!53}154.12&\cellcolor{black!33}17.82&\cellcolor{black!40}18.03&\cellcolor{black!20}3.91&\cellcolor{black!13}1.60&\cellcolor{black!6}0.29\\
        \hline
        \multirow{2}{*}{AIShell}&SI-SNR&$\cellcolor{black!0}\infty$&\cellcolor{black!33}26.02&\cellcolor{black!60}18.98&$-$&\cellcolor{black!53}19.22&\cellcolor{black!46}21.99&\cellcolor{black!40}25.99&\cellcolor{black!26}32.47&\cellcolor{black!20}32.54&\cellcolor{black!13}37.32 &\cellcolor{black!6}42.47\\
        &MSE&$\cellcolor{black!0}0.00$&\cellcolor{black!30}12.01&\cellcolor{black!52}58.86&$-$&\cellcolor{black!60}91.06&\cellcolor{black!45}39.99&\cellcolor{black!37}12.11&\cellcolor{black!22}2.73&\cellcolor{black!22}2.73&\cellcolor{black!15}0.92&\cellcolor{black!7}0.28\\
        
        \Xhline{1px}
        \end{tabular}
    \vspace{0.3cm}
    \end{subtable}

    \begin{subtable}[h]{\linewidth}
    \caption{EERs(\%) obtained in white- and black-box ASV evaluations. Results for the VoxCeleb dataset are presented for Vox-O, Vox-H, and Vox-E. Results for the AIShell dataset are presented in male (m) and female (f), respectively.}
        \centering
        \begin{tabular}{c | c | cc | ccccc | cc c | c}
        \Xhline{1px}
        \multirow{3}{*}{}&\multirow{3}{*}{Trials}&\multirow{3}{*}{Ori}&\multirow{3}{*}{Adv} & \multicolumn{5}{c|}{Ignorant}&\multicolumn{3}{c|}{\multirow{2}{*}{Semi-informed}}&\multirow{3}{*}{Well-informed}\\
        \cmidrule(r){5-9}
        &&&&\multicolumn{4}{c}{Non-informed Defense}& Denoising &\multicolumn{3}{c|}{}&\\
        \cmidrule(r){5-8}\cmidrule(r){9-9}\cmidrule(r){10-11} \cmidrule(r){12-12}
        &&&&QT &AAC& MS& AN&FRCRN-N&DCCRN-A&FRCRN-A&Denoising-G&\\
       \Xhline{1px}
        Copus&\multicolumn{11}{c}{White-box}\\
        \hline
        \multirow{3}{*}{VoxCeleb}&Vox-O&\cellcolor{black!9}1.46&\cellcolor{black!56}17.21&\cellcolor{black!30}4.80&\cellcolor{black!41}11.60&\cellcolor{black!43}12.04&\cellcolor{black!35}7.70&\cellcolor{black!54}17.20&\cellcolor{black!16}1.75&\cellcolor{black!15}1.70& \cellcolor{black!13}1.62&\cellcolor{black!11}1.47\\
        &Vox-H&\cellcolor{black!18}1.96&\cellcolor{black!58}21.56&\cellcolor{black!31}7.30&\cellcolor{black!46}16.67&\cellcolor{black!48}17.02&\cellcolor{black!37}11.07&\cellcolor{black!60}21.61&\cellcolor{black!26}2.52&\cellcolor{black!24}2.30& \cellcolor{black!22}2.16&\cellcolor{black!20}1.99\\
        &Vox-E&\cellcolor{black!0}1.05&\cellcolor{black!50}17.09&\cellcolor{black!28}4.36&\cellcolor{black!39}11.53&\cellcolor{black!45}12.48&\cellcolor{black!33}7.33&\cellcolor{black!52}17.12&\cellcolor{black!7}1.34&\cellcolor{black!5}1.23& \cellcolor{black!3}1.16&\cellcolor{black!1}1.08\\
        \hline
        \multirow{2}{*}{AIShell}&AIShell&\cellcolor{black!0}1.22&\cellcolor{black!45}34.25&\cellcolor{black!28}20.38&\cellcolor{black!37}32.78&\cellcolor{black!40}32.93&\cellcolor{black!34}26.33&\cellcolor{black!42}34.13&\cellcolor{black!20}4.84&\cellcolor{black!11}3.19&\cellcolor{black!8}2.89 &\cellcolor{black!2}1.67\\
        &AIShell-f&\cellcolor{black!5}2.87&\cellcolor{black!57}41.92&\cellcolor{black!31}25.59&\cellcolor{black!51}38.76&\cellcolor{black!54}39.28&\cellcolor{black!48}36.29&\cellcolor{black!60}42.04&\cellcolor{black!25}9.38&\cellcolor{black!17}4.11& \cellcolor{black!22}5.79&\cellcolor{black!14}3.43\\
        \Xhline{1px}
        Corpus&\multicolumn{11}{c}{Black-box}\\
        \hline        \multirow{3}{*}{VoxCeleb}&Vox-O&\cellcolor{black!3}1.69&\cellcolor{black!58}17.04&\cellcolor{black!32}7.76&\cellcolor{black!36}10.27&\cellcolor{black!38}10.45&\cellcolor{black!29}6.43&\cellcolor{black!48}12.65&\cellcolor{black!15}1.96&\cellcolor{black!13}1.95&\cellcolor{black!9}1.77 &\cellcolor{black!5}1.70\\
        &Vox-H&\cellcolor{black!17}2.87&\cellcolor{black!46}12.59&\cellcolor{black!44}11.63&\cellcolor{black!56}15.23&\cellcolor{black!54}15.19&\cellcolor{black!27}5.01&\cellcolor{black!60}17.11&\cellcolor{black!25}3.45&\cellcolor{black!23}3.28& \cellcolor{black!21}3.06&\cellcolor{black!19}2.91\\
        &Vox-E&\cellcolor{black!0}1.65&\cellcolor{black!50}13.01&\cellcolor{black!34}8.33&\cellcolor{black!40}11.01&\cellcolor{black!42}11.22&\cellcolor{black!30}6.60&\cellcolor{black!52}13.08&\cellcolor{black!15}1.96&\cellcolor{black!11}1.87& \cellcolor{black!7}1.73&\cellcolor{black!1}1.67\\
        \hline
        \multirow{2}{*}{AIShell}&AIShell-m&\cellcolor{black!0}1.16&\cellcolor{black!40}28.99&\cellcolor{black!34}27.14&\cellcolor{black!37}27.62&\cellcolor{black!45}29.31&\cellcolor{black!28}16.85&\cellcolor{black!42}29.01&\cellcolor{black!11}3.94&\cellcolor{black!8}2.48&\cellcolor{black!5}2.40 &\cellcolor{black!2}1.64\\
        &AIShell-f&\cellcolor{black!14}4.03&\cellcolor{black!57}38.64&\cellcolor{black!48}32.74&\cellcolor{black!54}38.18&\cellcolor{black!51}34.57&\cellcolor{black!31}26.10&\cellcolor{black!60}38.76&\cellcolor{black!25}12.07&\cellcolor{black!20}5.71& \cellcolor{black!22}6.95&\cellcolor{black!17}5.50\\
        \Xhline{1px}
        \end{tabular}
        \vspace{0.3cm}
        \end{subtable}

        

    \begin{subtable}[t]{\linewidth}
    \centering
    \caption{Utility evaluation results. PESQ, WER, Pitch-Mean, and Pitch-STD values are included.}
    \begin{tabular}{c | c | cc | ccccc | ccc | c}
        \Xhline{1px}
        \multirow{3}{*}{Corpus}&\multirow{3}{*}{}&\multirow{3}{*}{Ori}&\multirow{3}{*}{Adv} & \multicolumn{5}{c|}{Ignorant}&\multicolumn{3}{c|}{\multirow{2}{*}{Semi-informed}}&\multirow{3}{*}{Well-informed}\\
        \cmidrule(r){5-9}
        &&&&\multicolumn{4}{c}{Non-informed Defense}& Denoising &\multicolumn{3}{c|}{}&\\
        \cmidrule(r){5-8}\cmidrule(r){9-9}\cmidrule(r){10-11} \cmidrule(r){12-12}
        &&&&QT &AAC& MS& AN&FRCRN-N&DCCRN-A&FRCRN-A&Denoising-G&\\
       \Xhline{1px}
        \multirow{4}{*}{VoxCeleb}&PESQ&\cellcolor{black!0}4.50&\cellcolor{black!33}3.72&\cellcolor{black!53}3.34&\cellcolor{black!60}3.16&\cellcolor{black!40}3.48&\cellcolor{black!46}3.42&\cellcolor{black!33}3.72&\cellcolor{black!26}4.17&\cellcolor{black!20}4.32&\cellcolor{black!13}4.40 &\cellcolor{black!6}4.47\\
        &Pitch-Mean&\cellcolor{black!0}1.00&\cellcolor{black!24}0.98&\cellcolor{black!36}0.96&\cellcolor{black!60}0.87&\cellcolor{black!36}0.96&\cellcolor{black!36}0.96&\cellcolor{black!12}0.99&\cellcolor{black!48}0.93&\cellcolor{black!12}0.99&\cellcolor{black!0}1.00 &\cellcolor{black!0}1.00\\\
        &Pitch-STD&\cellcolor{black!0}0.00&\cellcolor{black!30}0.04&\cellcolor{black!50}0.07&\cellcolor{black!60}0.08&\cellcolor{black!40}0.06&\cellcolor{black!50}0.07&\cellcolor{black!30}0.04&\cellcolor{black!40}0.06&\cellcolor{black!30}0.04& \cellcolor{black!20}0.02&\cellcolor{black!10}0.01\\
        \hline
        \multirow{4}{*}{AIShell}&PESQ&\cellcolor{black!0}4.50&\cellcolor{black!33}3.19&\cellcolor{black!60}2.47&\cellcolor{black!46}2.99&\cellcolor{black!40}3.12&\cellcolor{black!53}2.93&\cellcolor{black!33}3.19&\cellcolor{black!26}3.72&\cellcolor{black!20}4.07& \cellcolor{black!13}4.17&\cellcolor{black!6}4.33\\
        
        &Pitch-Mean&\cellcolor{black!0}1.00&\cellcolor{black!37}0.80&\cellcolor{black!45}0.79&\cellcolor{black!52}0.75&\cellcolor{black!30}0.82&\cellcolor{black!60}0.65&\cellcolor{black!37}0.80&\cellcolor{black!22}0.93&\cellcolor{black!15}0.96& \cellcolor{black!7}0.98&\cellcolor{black!7}0.98\\
        &Pitch-STD&\cellcolor{black!0}0.00&\cellcolor{black!52}0.26&\cellcolor{black!37}0.23&\cellcolor{black!45}0.24&\cellcolor{black!30}0.21&\cellcolor{black!60}0.34&\cellcolor{black!52}0.26&\cellcolor{black!22}0.09&\cellcolor{black!15}0.06& \cellcolor{black!7}0.04&\cellcolor{black!15}0.06\\
        \Xhline{1px}
    \end{tabular}
    \end{subtable}

    \end{table*}

Next, the ignorant, semi-informed, and well-informed evaluations were performed on cross-domain datasets, including VoxCeleb\cite{nagrani2017voxceleb} and AIShell-1\cite{aishell_2017}. The speaker-adversarial speech was generated with the SSED generator. Specifically, VoxCeleb primarily consisted of English utterances, similar to LibriSpeech, with which the perturbation generator and remover were trained. The official test trials of VoxCeleb were evaluated, including Vox-O, Vox-H, and Vox-E. In contrast, AIShell is a Chinese dataset, which is a language distinct from the training data. The official test set was applied for evaluation, which contains 20 speakers. For each speaker in the AIShell test set, 1 pair of target trials and 30 pairs of non-target trials were constructed in a gender-dependent manner. Utility evaluations were carried out on all utterances within their respective test sets. Due to the absence of text transcriptions in the VoxCeleb data, intelligibility evaluations were skipped.

The results are given in Table \ref{tb. cross domain}. From Table \ref{tb. cross domain}, it can be observed that the results on VoxCeleb are consistent with those obtained on the LibriSpeech evaluation datasets, as presented in Tables \ref{tb. ignorant} - \ref{tb. well-informed}. Specifically, in the ignorant evaluations on the VoxCeleb datasets, the non-informed defense methods effectively reduced the impacts of the perturbations on speaker attribute extraction, as evidenced by the lower EERs in the ASV evaluations compared to the adversarial speech. However, they were unable to eliminate the impacts of the perturbations, indicated by the higher EERs than the recordings. Moreover, the processed speech exhibited higher MSE and Pitch-STD values, lower SI-SNR, PESQ, and Pitch-Mean values than the adversarial speech, implying adverse impacts on original speech restoration and utility. Besides, the FRCRN-N model failed to reduce EERs in ASV evaluations when applied to adversarial speech. In the speech restoration evaluation, the FRCRN-N model was ineffective in reducing perturbations, as indicated by the comparable SI-SNR and MSE values obtained on the processed and adversarial speech. In utility evaluations, the FRCRN-N model failed to improve speech utility, as indicated by PESQ, Pitch-Mean, and Pitch-STD values comparable to those of adversarial speech. The inability to remove perturbations was consistently observed in the AIShell dataset.

In the semi-informed evaluations, the FRCRN-N model effectively reduced the impact of perturbations on speaker attribute extraction across both the VoxCeleb and AIShell datasets. However, it demonstrated unstable performance in original speech restoration and utility. The FRCRN-A model was effective in reducing the perturbations as indicated by the reduced MSE and increased SI-SNR compared with the adversarial speech. Besides, it was able to mitigate the impact of adversarial perturbations on speaker attribute extraction as implied by the lower EERs in the ASV evaluations than the adversarial speech. Moreover, it improved the utility of the utterances, supported by the higher PESQ, SI-SNR, and Pitch-Mean values and lower Pitch-STD values than the adversarial speech. The capability of perturbation removal was further enhanced in the Denoising-G method. Overall, the perturbation removal capability in the semi-informed scenario on both the VoxCeleb and AIShell datasets was consistent with that observed on the in-domain LibriSpeech test subset as presented previously.

In the well-informed evaluations, the processed speech achieved the lowest EERs, highest PESQ, SI-SNR, and Pitch-Mean values, and the lowest Pitch-STD values among the three evaluation scenarios, suggesting its superior perturbation removal capability. Specifically, for the VoxCeleb utterances, an SI-SNR of 47.81 (approaching 50) and an MSE of 0.29, demonstrated the capability of almost removing the perturbations. Besides, the ASV evaluations of the white-box models yielded comparable EERs to those of the recordings. Moreover, a PESQ of 4.47 (near optimal 4.50), a Pitch-Mean of 1.00, and a Pitch-STD of 0.01 were achieved, collectively indicating near-complete restoration of the utility of the recording. In the AIShell dataset, which was in a different language from the training data, the restoration effectiveness was slightly reduced, as implied by the slightly higher EERs in the white-box ASV evaluations, a lower SI-SNR value of 42.47, a PESQ value of 4.33, further from the optimal 4.50; further from 50; a Pitch-Mean value of 0.98, marginally below 1.00; and a Pitch-STD value of 0.06, slightly above 0.01. This indicated a marginal degradation in the perturbation removal capability on the cross-language dataset.

\begin{table}[t]
    \caption{Comparison between independently and jointly trained generators and removers. Given the jointly trained generator, the results from both the semi-informed remover (Denoising-G) and the jointly trained remover are included. \emph{Ori} and \emph{Adv} are short for original and adversarial speech, respectively.}
    \label{tb. ablation study}
    \centering
    \scriptsize

    \begin{subtable}[t]{\linewidth}
    \centering
    \caption{Original speech restoration evaluation results. SI-SNR and MSE values are included.}
    \begin{tabular}{ c | c | c c c | c c}
    \Xhline{1px}

    &\multirow{2}{*}{Ori}&\multicolumn{3}{c|}{Joint}&\multicolumn{2}{c}{Independent}\\
    \cmidrule(r){3-7}
    &&Adv&Denoising-G&Remover&Adv&Remover\\
    \Xhline{1px}    SI-SNR&$\infty$&32.43&   42.46  &\textbf{50.29}&33.19&36.41\\
    MSE&0&17.45&   1.80    &\textbf{0.28}&14.72&12.97\\
    \Xhline{1px}
    \end{tabular}
    \vspace{0.3cm}
    \end{subtable}

    \begin{subtable}[t]{\linewidth}
    \centering
    \caption{EERs(\%) obtained in white- and black-box ASV evaluations. Results are presented on libri-m and libri-f, respectively.}
    \begin{tabular}{ c | c | c c c | c c}
    \Xhline{1px}

    \multirow{2}{*}{Trials}&\multirow{2}{*}{Ori}&\multicolumn{3}{c|}{Joint}&\multicolumn{2}{c}{Independent}\\
    \cmidrule(r){3-7}
    &&Adv&Denoising-G&Remover&Adv&Remover\\
    \Xhline{1px}
     \multicolumn{7}{c}{White-box}\\ 
    \hline
    libri-m&1.21&18.62& 1.21   &\textbf{1.21}&23.99&1.53\\
    libri-f&1.64&19.27&   1.99  &\textbf{1.64}&22.14&2.29\\
    \Xhline{1px}
     \multicolumn{7}{c}{Black-box}\\ 
    \hline
    libri-m&1.21&12.17&  1.37   &\textbf{1.15}&14.57&1.31\\
    libri-f&2.23&11.78&   2.52  &\textbf{2.29}&14.00&3.16\\
    \Xhline{1px}
    \end{tabular}
    \vspace{0.3cm}
    \end{subtable}

    \begin{subtable}[t]{\linewidth}
    \centering
    \caption{Utility evaluation results. PESQ, WER, Pitch-Mean, and Pitch-STD values are included.}
    \begin{tabular}{c | c | c c c | c c}
    \Xhline{1px}
   &\multirow{2}{*}{Ori}&\multicolumn{3}{c|}{Joint}&\multicolumn{2}{c}{Independent}\\
    \cmidrule(r){3-7}
    &&Adv&Denoising-G&Remover&Adv&Remover\\
    \Xhline{1px}
    PESQ&4.50&3.67& 4.36   &\textbf{4.47}&3.81&4.13\\
    WER&4.08&6.94&  4.12  &\textbf{4.08}&5.99&4.29\\
    Pitch-Mean&1.00&0.99&  1.00  &1.00&0.99&0.99\\
    Pitch-STD&0&0.03&  0.01  &0.01&0.03&0.02\\
    \Xhline{1px}
    \end{tabular}
    \end{subtable}
\end{table}

\subsection{Ablation study}
Finally, experiments were conducted for an ablation study to examine the impacts of the joint training mechanism in perturbation removal within the well-informed scenario. To this end, a speaker-adversarial speech generator and remover, both based on the SSED architecture, were trained independently. That is, the perturbation generator was trained first. Then, the adversarial versions of the training utterances were generated and used to train a perturbation remover. Besides, the generator jointly trained with the remover was examined. On the adversarial speech generated by the jointly trained generator, two removers were evaluated for comparison: 1) the remover independently trained with pairs of original and adversarial speech, specified as the Denoising-G remover examined in the semi-informed evaluation as presented in Table \ref{tb. semi-informed}, and 2) the jointly trained remover as presented in Table \ref{tb. well-informed}. The awareness of the generator by the remover increased gradually across the three configurations, specified as follows. In the independent training, the generator and remover were unaware of the internal attributes of each other. In the Denoising-G configuration, the generator was trained with constraints from a remover sharing the same architecture as the Denoising-G remover, making it aware of the architecture of the remover. In the joint training configuration, both the remover and generator were aware of the internal attributes of each other.

Table \ref{tb. ablation study} presents the results of the three configurations. The two generators were trained to generate adversarial speech with comparable SI-SNR values for a fair comparison. As shown in the table, the independently trained remover achieved significantly lower EERs than adversarial speech in both white- and black-box ASV evaluations, demonstrating its effectiveness in reducing the impact of the speaker-adversarial perturbations on speaker attribute extraction. However, the EERs were higher than those obtained from the original recordings. Additionally, it failed to eliminate perturbations, as indicated by an SI-SNR of 36.41 (lower than 50) and an MSE of 12.97 (higher than 0). Moreover, a PESQ of 4.13, and a WER of 4.29 further demonstrated its incapability to restore the utility of the original speech. These results demonstrate the incapability of the independently trained remover in restoring original recordings. Comparison with the jointly trained remover, which nearly eliminated the perturbations, demonstrated the effectiveness of the joint training mechanism in perturbation removal. Lastly, the Denoising-G remover demonstrated performances that were between those of the independently and jointly trained removers across all evaluation metrics. This indicated that an increase in the awareness between the generator and remover led to improved perturbation elimination capability.

\section{Conclusions}


This paper studied the removability of speaker-adversarial perturbations in the ignorant, semi-informed, and well-informed scenarios. The study was conducted on the perturbations generated with both optimization-based and feedforward algorithms. The experimental findings obtained on the LibriSpeech dataset demonstrate three key findings. First, in the ignorant scenario, the speaker-adversarial perturbations cannot be eliminated. Second, in the semi-informed scenario, the speaker-adversarial perturbations cannot be fully removed, while those generated by the feedforward model can be considerably reduced. Third, in the well-informed scenario, speaker-adversarial perturbations are nearly eliminated, allowing for the restoration of the original speech.
    

\bibliographystyle{IEEEtran}
\bibliographystyle{unsrt}
\input{myref.bbl}

\end{document}